\documentclass[11pt,showpacs,showkeys]{revtex4-1}
\usepackage[utf8x]{inputenc}
\usepackage{ucs}
\usepackage{amsthm}
\usepackage{amsmath,amssymb}
\usepackage{latexsym}
\usepackage{amsfonts}
\usepackage{fancybox, calc}  
\usepackage{dcolumn}
\usepackage{bm}
\usepackage[usenames]{color}
\usepackage{multirow}
\usepackage{graphicx}
\usepackage{hyperref}
\usepackage{subfig}
\usepackage{booktabs}
\usepackage{xcolor}
\usepackage{makeidx}
\usepackage[normalem]{ulem}
\usepackage{float} 
\usepackage{wrapfig} 
\usepackage{upgreek} 
\usepackage{cancel} 
\usepackage{mathdots} 
\usepackage{mathrsfs} 
\graphicspath{{figures/}} 
\newtheorem{thm}{Theorem}

\def\dsp{\displaystyle}

\begin{document}

\title{ Entropic uncertainty measures for large dimensional \\ 
hydrogenic systems}

\author{D. Puertas-Centeno}
\email[]{vidda@correo.ugr.es}
\address{Departamento de F\'{\i}sica At\'{o}mica, Molecular y Nuclear, Universidad de Granada, Granada 18071, Spain\\ Instituto Carlos I de F\'{\i}sica Te\'orica y Computacional, Universidad de Granada, Granada 18071, Spain}

\author{N.M. Temme}
\email[]{Nico.Temme@cwi.nl}
\affiliation{Former address: CWI, Science Park 123, 1098 XG Amsterdam, The Netherlands\\ IAA, 1825 BD 25, Alkmaar, The Netherlands}

\author{I.V. Toranzo}
\email[]{ivtoranzo@ugr.es}
\affiliation{Departamento de F\'{\i}sica At\'{o}mica, Molecular y Nuclear, Universidad de Granada, Granada 18071, Spain}
\affiliation{Instituto Carlos I de F\'{\i}sica Te\'orica y Computacional, Universidad de Granada, Granada 18071, Spain}

\author{J.S. Dehesa}
\email[]{dehesa@ugr.es}
\affiliation{Departamento de F\'{\i}sica At\'{o}mica, Molecular y Nuclear, Universidad de Granada, Granada 18071, Spain}
\affiliation{Instituto Carlos I de F\'{\i}sica Te\'orica y Computacional, Universidad de Granada, Granada 18071, Spain}

\begin{abstract}
The entropic moments of the probability density of a quantum system in position and momentum spaces describe not only some fundamental and/or experimentally accessible quantities of the system, but also the entropic uncertainty measures of Rényi type which allow one to find the most relevant mathematical formalizations of the position-momentum Heisenberg's uncertainty principle, the entropic uncertainty relations. It is known that the solution of difficult three-dimensional problems can be very well approximated by a series development in $1/D$ in similar systems with a non-standard dimensionality $D$; moreover, several physical quantities of numerous atomic and molecular systems have been numerically shown to have values in the large-$D$ limit comparable to the corresponding ones provided by the three-dimensional numerical self-consistent field methods. The $D$-dimensional hydrogenic atom is the main prototype of the physics of multidimensional many-electron systems. In this work we rigorously determine the leading term of the Rényi entropies of the $D$-dimensional hydrogenic atom at the limit of large $D$. As a byproduct, we show that our results saturate the known position-momentum R\'enyi-entropy-based uncertainty relations. 
\end{abstract}



\maketitle

\section{Introduction}

In an excellent tutorial article about quarks, gluons and \textit{impossible problems} of quantum chromodynamics, Edward Witten \cite{witten} illustrated the utility of the large-dimension $D$ limit with a rough calculation for helium. This prompted Dudley R. Herschbach et al \cite{herschbach1993,tsipis} and other authors (see the review \cite{chatterjee}) to develop a new strategy, the dimensional scaling method, to solve first the quantum problems with one degree of freedom \cite{chatterjee} and later the much more difficult Coulomb problems involving two or more non-separable, strongly-coupled degrees of freedom which usually take place in physics of atoms and molecules \cite{herschbach1993,tsipis}. With this method a finite many-body problem is typically solved in the large-$D$ limit, most often in an analytical way, and then perturbation theory in $1/D$ is used to obtain an approximate result for the standard dimension ($D=3$).\\

Physics in the large-$D$ limit becomes much simpler. Indeed, in this limit the electrons of a many-electron system assume fixed positions relative to the nuclei and each other, in the $D$-scaled space. Moreover, the large-$D$ electronic geometry and energy correspond to the minimum of an exactly known effective potential and can be determined from classical electrostatics for any atom or molecule, what remembers the prequantum models of Lewis and Langmuir \cite{herschbach86,herschbach_2000}. The ($D\to\infty$)-limit is called \textit{pseudoclassical}, tantamount to $h\to 0$ and/or $m_e\to\infty$ in the kinetic energy. This limit is not the same as the conventional classical limit obtained by $h\to 0$ for a fixed dimension \cite{yaffe82,yaffe83}. Although at first sight the electrons at rest in fixed locations might seem violate the uncertainty principle, this is not true because that occurs only in the $D$-scaled space (see e.g., \cite{herschbach_2000}). For $D$ finite but very large, the electrons are confined to harmonic oscillations about the fixed positions attained in the ($D\rightarrow\infty$)-limit. Moreover, the large-$D$ limit of numerous physical properties of almost all atoms with up to $100$ electrons and many diatomic molecules have been numerically evaluated, obtaining values comparable to or better than single-zeta Hartree-Fock calculations \cite{herschbach1993,tsipis,herschbach96}.\\

The main prototype of the $D$-dimensional Coulomb many-body systems, the $D$-dimensional hydrogen atom (i.e., a negatively-charged particle moving in a space of $D$ dimensions around a positively charged core which electromagnetically binds it in its orbit), has been investigated in detail starting from its wave functions which are analytically known \cite{yanez_1994,nieto,dehesa_2010} in the two conjugated position and momentum spaces for any dimension. This system includes a wide variety of physical objects, such as e.g. hydrogenic atoms and ions, some exotic atoms and antimatter atoms, excitons in semiconductors and qubits.\\

The spreading properties of the electronic distribution of the $D$-dimensional hydrogenic atom have been analyzed by means of its moments around the origin (radial expectation values) in both position \cite{pasternack,ray,andrae,drake,tarasov,guerrero11} and momentum \cite{hey2,assche} spaces. However, these quantities are formally given in terms of $D$, the hyperquantum numbers of the hydrogenic states and the nuclear charge $Z$ through a generalized hypergeometric function $_{p+1}F_{p}(1)$, which cannot be easily calculated unless the hyperquantum numbers and/or the dimension $D$ are sufficiently small. Recently the position and momentum moments around the origin  of the $D$-dimensional hydrogenic atom have been determined in a simple and compact form for the highly and very-highly excited (i.e., Rydberg) states \cite{dehesa_2010,aptekarev} as well as for any excited state at large $D$ \cite{tor2016}.\\ 

The determination of the entropic measures of the $D$-dimensional hydrogenic atom, which describe most appropriately the electronic uncertainty of the system, is far more difficult except for the lowest-lying energy states despite some efforts \cite{dehesa_2010}. This is because these quantities are described by means of some power or logarithmic functionals of the electron density, which cannot be calculated in an analytical way nor numerically computed; the latter is basically because a naive numerical evaluation using quadratures is not convenient due to the increasing number of integrable singularities when the principal hyperquantum number $n$ is increasing, which spoils any attempt to achieve reasonable accuracy even for rather small $n$ \cite{buyarov}. Recently, the main entropic properties of the $D$-dimensional Rydberg hydrogenic states (namely, the Rényi, Shannon and Tsallis entropies) have been explicitly calculated in a compact form \cite{toranzo16a,toranzo16b} by use of modern techniques of approximation theory based on the strong asymptotics ($n\to\infty$) of the Laguerre $\mathcal{L}^{(\alpha)}_{n}(x)$ and Gegenbauer  $\mathcal{C}^{(\alpha)}_{n}(x)$ polynomials which control the state's wave functions in position and momentum spaces, respectively \cite{aptekarev2016}.\\

In this work we first determine the R\'enyi entropy and then we conjecture the Shannon entropy in both position and momentum spaces for the large-dimensional hydrogenic states in terms of the dimensionality $D$, the nuclear charge $Z$ and the principal and orbital hyperquantum numbers of the states. The Rényi entropies $R_{q}[\rho], q>0$ are defined \cite{renyi1,shannon} as
\begin{equation}
\label{eq:renentropa}
R_{q}[\rho] =  \frac{1}{1-q}\log \int_{\mathbb{R}^D} [\rho(\vec{r})]^{q}\, d\vec{r},\quad q\neq1.
\end{equation}  
Note that the Shannon entropy $S[\rho] = - \int \rho(\vec{r}) \log \rho(\vec{r}) d\vec{r} = \lim_{q\rightarrow 1} R_{q}[\rho]$; see e.g. \cite{aczel}. These quantities completely characterize the density $\rho(\vec{r})$ \cite{romera_01} under certain conditions. In fact, we can calculate from (\ref{eq:renentropa}) other relevant entropic quantities such as e.g. the disequilibrium $\langle\rho\rangle = \exp(-R_{2}[\rho])$, and the Tsallis entropies $T_{q}[\rho] =  \frac{1}{q-1} \left(1-\int_{\mathbb{R}^D} [\rho(\vec{r})]^{q}\,d\vec r\right),q >0$ \cite{tsallis} as
\begin{eqnarray}
\label{eq:rel2}
T_{q}[\rho] &=& \frac{1}{1-q}[e^{(1-q)R_{q}[\rho]}-1].
\end{eqnarray}
which holds for $q\neq1$. Here again, the Shannon entropy $S[\rho] = \lim_{q\rightarrow 1} T_{q}[\rho]$.
The properties of the Rényi entropies and their applications have been widely considered; see e.g. \cite{aczel,leonenko,jizba_2004b} and the reviews \cite{dehesa_sen12,bialynicki3,jizba}. The use of Rényi and Shannon entropies as measures of uncertainty allow a wider quantitative range of applicability than the moments around the origin and the standard or root-square-mean deviation do. This permits, for example, a quantitative discussion of quantum uncertainty relations further beyond the conventional Heisenberg-like uncertainty relations \cite{hall,dehesa_sen12,bialynicki3,tor2016}.


The structure of this work is the following. In section \ref{sec:basics} the wave functions of the $D$-dimensional hydrogenic states in both position and momentum spaces are briefly described, and the corresponding probability densities are given. In section \ref{sec:renyi} we determine the physical Rényi entropies of the $D$-dimensional hydrogenic atom at large $D$ by use of some recent theorems relative to the asymptotics ($\alpha \to \infty$) of the underlying Rényi-like integral functionals of Laguerre polynomials $\mathcal{L}^{(\alpha)}_{k}(x)$ and Gegenbauer polynomials $\mathcal{C}^{(\alpha)}_{k}(x)$ which control the hydrogenic wavefunctions as described in the previous section. The dominant term of the joint position-momentum uncertainty sum for the general states of the large dimensional hydrogenic systems is also given and, what is most interesting, shown to saturate the known position-momentum R\'enyi-entropy-based uncertainty relations \cite{bialynicki2,vignat,zozor2008}. Finally, some conclusions, open problems and three appendices are given.

\section{The $D$-dimensional hydrogenic problem: Basics}
\label{sec:basics}
In this section we briefly summarize the physical solutions of the Schr\"{o}dinger equation of the $D$-dimensional hydrogenic system in both position and momentum spaces. Then we give the associated position and momentum $D$-dimensional probability densities of the system.\\

The time-independent Schr\"{o}dinger equation of a $D$-dimensional ($D >1$) hydrogenic system (i.e., an electron moving under the action of the $D$-dimensional Coulomb potential $\displaystyle{V(\vec{r})=-\frac{Z}{r}}$) is given by
\begin{equation}\label{eqI_cap1:ec_schrodinger}
\left( -\frac{1}{2} \vec{\nabla}^{2}_{D} - \frac{Z}{r}
\right) \Psi \left( \vec{r} \right) = E \Psi \left(\vec{r} \right),
\end{equation}
where $\vec{\nabla}_{D}$ denotes the $D$-dimensional gradient operator, $Z$ is the nuclear charge, and the  electronic position vector $\vec{r}  =  (x_1 ,  \ldots  , x_D)$ in hyperspherical units  is  given as $(r,\theta_1,\theta_2,\ldots,\theta_{D-1})      \equiv
(r,\Omega_{D-1})$, $\Omega_{D-1}\in S^{D-1}$, where $r \equiv |\vec{r}| = \sqrt{\sum_{i=1}^D x_i^2}
\in [0 , \: +\infty)$  and $x_i =  r \left(\prod_{k=1}^{i-1}  \sin \theta_k
\right) \cos \theta_i$ for $1 \le i \le D$
and with $\theta_i \in [0 , \pi), i < D-1$, $\theta_{D-1} \equiv \phi \in [0 ,  2
\pi)$. It is assumed that the nucleus is located at the origin and, by  convention, $\theta_D =  0$ and the  empty product is the  unity. Atomic units are used throughout the paper. \\
It is known \cite{nieto,yanez_1994,dehesa_2010} that the energies belonging to the discrete spectrum are given by  
\begin{equation} \label{eqI_cap1:energia}
E= -\frac{Z^2}{2\eta^2},\hspace{0.5cm} \hspace{0.5cm} \eta=n+\frac{D-3}{2}; \hspace{5mm} n=1,2,3,\ldots,
\end{equation}
and the associated eigenfunction can be expressed as
\begin{equation}\label{eqI_cap1:FunOnda_P}
\Psi_{n,l, \left\lbrace \mu \right\rbrace }(\vec{r})=\mathcal{R}_{n,l}(r)\,\, {\cal{Y}}_{l,\{\mu\}}(\Omega_{D-1}),
\end{equation}
where $(l,\left\lbrace \mu \right\rbrace)\equiv(l\equiv\mu_1,\mu_2,\ldots,\mu_{D-1})$ denote the hyperquantum numbers associated to the angular variables $\Omega_{D-1}\equiv (\theta_1, \theta_2,\ldots,\theta_{D-1})$, which may take all values consistent with the inequalities $l\equiv\mu_1\geq\mu_2\geq\ldots\geq \left|\mu_{D-1} \right| \equiv \left|m\right|\geq 0$. The radial eigenfunction is given by
\begin{align}\label{eqI_cap1:Rnl}
\mathcal{R}_{n,l}(r)&=\mathfrak N_{n,l}\left(\frac{r}{\lambda}\right)^{l}e^{-\frac{r}{2\lambda}} \mathcal{L}_{n-l-1}^{(2l+D-2)}\left(\frac{r}{\lambda}\right)\\ \nonumber
&=\mathfrak N_{n,l}\left[\frac{\omega_{2L+1}(\tilde{r})}{\tilde{r}^{D-2}}\right]^{1/2}{\cal{L}}_{\eta-L-1}^{(2L+1)}(\tilde{r})\\ \nonumber
&= \left( \frac{\lambda^{-D}}{2 \eta}\right)^{1/2}   \left[\frac{\omega_{2L+1}(\tilde{r})}{\tilde{r}^{D-2}}\right]^{1/2}{\widehat{\cal{L}}}_{\eta-L-1}^{(2L+1)}(\tilde{r}),
\end{align}
where the ``grand orbital angular momentum quantum number'' $L$ and the dimensionless parameter $\tilde{r}$ are
\begin{align} \label{eqI_cap1:Lyr}
L&=l+\frac{D-3}{2}, \hspace{0.5cm} l=0, 1, 2, \ldots \\ \label{rtilde}
\tilde{r}&=\frac{r}{\lambda},\hspace{0.5cm} 
\hspace{0.5cm}\lambda=\frac{\eta}{2Z},
\end{align}
and
$\omega_{\alpha}(x) =x^{\alpha}e^{-x}, \, \alpha= 2L +1 = 2l+D-2,$
  is the weight function of the Laguerre polynomials $\mathcal{L}_{k}^{(\alpha)}(x), x \in \left[0,\infty \right)$. Note that $\alpha \geq 0$ for $D \geq 2$. The symbols $\mathcal{L}_{k}^{(\alpha')}(x)$ and $\widehat{\mathcal{L}}_{k}^{(\alpha')}(x)$ denote the orthogonal and orthonormal Laguerre polynomials, so that
\begin{equation}\label{eqI_cap1:laguerre_orto_ortogo}
{\widehat{\mathcal{L}}}^{(\alpha')}_{k}(x)=  \left( \frac{k!}{\Gamma(k+\alpha'+1)}\right)^{1/2}  {\mathcal{L}}^{(\alpha')}_{k}(x),
\end{equation}
for any parameter $\alpha' > -1$, and finally
\begin{equation}
\mathfrak N_{n,l}  \equiv \lambda^{-\frac{D}{2}}\left\{\frac{(\eta-L-1)!}{2\eta(\eta+L)!}\right\}^{\frac{1}{2}}=\left\{\left(\frac{2Z}{n+\frac{D-3}{2}}\right)^{D}\frac{(n-l-1)!}{2\left(n+\frac{D-3}{2}\right)(n+l+D-3)!}  \right\}^{\frac{1}{2}}
\label{eq:4}
\end{equation}
represents the normalization constant which ensures that $\int \left| \Psi_{\eta,l, \left\lbrace \mu \right\rbrace }(\vec{r}) \right|^2 d\vec{r} =1$. Note that the $D$-dimensional volume element is $d\vec{r} \equiv d^{D}r = r^{D-1}\,dr\, d\Omega_{D-1}$ and
\[
d\Omega_{D-1}=\left(\prod_{j=1}^{D-2} (\sin \theta_j)^{2 \alpha_j} d\theta_j\right) d\theta_{D-1}.
\]
where $2\alpha_{j} = D-j-1$. The angular eigenfunctions are the hyperspherical harmonics, $\mathcal{Y}_{l,\{\mu\}}(\Omega_{D-1})$, defined \cite{yanez_1994,dehesa_2010,avery} as
\begin{equation} 
\mathcal{Y}_{l,\{\mu\}}(\Omega_{D-1}) = \mathcal{N}_{l,\{\mu\}}e^{im\phi}\times \prod_{j=1}^{D-2}\mathcal{C}^{(\alpha_{j}+\mu_{j+1})}_{\mu_{j}-\mu_{j+1}}(\cos\theta_{j})(\sin\theta_{j})^{\mu_{j+1}}
\label{eq:hyperspherarm}
\end{equation}
with the squared normalization constant given as
\begin{eqnarray}
\label{eq:normhypersphar}
\mathcal{N}_{l,\{\mu\}}^{2} &=& \frac{1}{2\pi}
\prod_{j=1}^{D-2} \frac{(\alpha_{j}+\mu_{j})(\mu_{j}-\mu_{j+1})![\Gamma(\alpha_{j}+\mu_{j+1})]^{2}}{\pi \, 2^{1-2\alpha_{j}-2\mu_{j+1}}\Gamma(2\alpha_{j}+\mu_{j}+\mu_{j+1})},\nonumber\\
&=& \frac{1}{2\pi}\prod_{j=1}^{D-2}A_{\mu_{j},\mu_{j+1}} ^{(j)},
\end{eqnarray}
where the symbol $\mathcal{C}^{(\alpha')}_{k}(t)$ denotes the Gegenbauer polynomial \cite{nist1} of degree $k$ and parameter $\alpha'$.\\
Then, the quantum probability density of a $D$-dimensional hydrogenic stationary state $(n,l,\{\mu\})$ is given in position space by the the squared modulus of the position eigenfunction given by (\ref{eqI_cap1:FunOnda_P}) as 
\begin{equation}
\label{eq:denspos}
\rho_{n,l,\{\mu\}}(\vec{r}) = \rho_{n,l}(\tilde{r})\,\, |\mathcal{Y}_{l,\{\mu\}}(\Omega_{D-1})|^{2},
\end{equation}
where the radial part of the density is the univariate radial density function
\begin{eqnarray}
\label{eq:radensity}
\rho_{n,l}(\tilde{r}) &=& [\mathcal{R}_{n,l}(r)]^2 = \frac{\lambda^{-D}}{2 \eta} \,\, \frac{\omega_{2L+1}(\tilde{r})}{\tilde{r}^{D-2}}\,\,[{\widehat{\mathcal{L}}}_{\eta-L-1}^{(2L+1)}(\tilde{r})]^{2}.
\end{eqnarray}
On the other hand, the Fourier transform of the position eigenfunction $\Psi_{\eta,l, \left\lbrace \mu \right\rbrace }(\vec{r})$ given by (\ref{eqI_cap1:FunOnda_P}) provides the eigenfunction of the system in the conjugated momentum space as
\begin{equation}
\label{eq:momwvf}
\tilde{\Psi}_{n,l,\{\mu\}}(\vec{p}) = \mathcal{M}_{n,l}(p)\,\,\mathcal{Y}_{l,\{\mu \}}(\Omega_{D-1}),
\end{equation}
where the radial part is 
\begin{equation}
\label{eq:radmomwvf}
\mathcal{M}_{n,l}(p) = K_{n,l}\frac{(\eta \tilde{p})^{l}}{(1+\eta^{2}\tilde{p}^{2})^{L+2}}\mathcal{C}_{\eta-L-1}^{(L+1)}\left( \frac{1-\eta^{2}\tilde{p}^{2}}{1+\eta^{2}\tilde{p}^{2}} \right)
\end{equation}
with $\tilde{p} =\frac{p}{Z}$ and the normalization constant
\begin{equation}
\label{eq:normacons}
K_{n,l} = Z^{-\frac{D}{2}}2^{2L+3}\left[\frac{(\eta-L-1)!}{2\pi(\eta+L)!} \right]^{\frac{1}{2}}\Gamma(L+1)\eta^{\frac{D+1}{2}}.
\end{equation}
Then, the expression  
\begin{eqnarray}\label{eq:momdens}
\gamma_{n,l,\{\mu\}}(\vec{p}) &=& |\tilde{\Psi}_{n,l,\{\mu \}}(\vec{p})|^{2} =  \mathcal{M}^{2}_{n,l}(p) \,\,|\mathcal{Y}_{l,\{\mu \}}(\Omega_{D-1})|^{2}\nonumber \\
&=& K_{n,l}^{2}\frac{(\eta \tilde{p})^{2l}}{(1+\eta^{2}\tilde{p}^{2})^{2L+4}}\left[\mathcal{C}_{\eta-L-1}^{(L+1)}\left( \frac{1-\eta^{2}\tilde{p}^{2}}{1+\eta^{2}\tilde{p}^{2}} \right)\right]^{2}|\mathcal{Y}_{l,\{\mu \}}(\Omega_{D-1})|^{2}
\end{eqnarray}
gives the momentum probability density of the $D$-dimensional hydrogenic stationary state with the hyperquantum numbers $(n,l,\{\mu\})$.

\section{Rényi entropies of large dimensional hydrogenic states}
\label{sec:renyi}
In this section we obtain the Rényi entropies of a generic $D$-dimensional hydrogenic state $(n,l,\{\mu\})$ in the large-$D$ limit in both position and momentum spaces. We start with the expressions (\ref{eq:denspos}) and (\ref{eq:momdens}) of the position and momentum probability densities of the system, respectively.\\

 To calculate the position Rényi entropy we decompose it into two radial and angular parts. The radial part is first expressed in terms of a Rényi-like integral functional of Laguerre polynomials $\mathcal{L}_{n-l-1}^{(\alpha)}(x)$ with $\alpha = D +2l-2$, and then this functional is determined in the large-$D$ limit by means of Theorem \ref{T1bis} (see Appendix \ref{A}). The angular part is given by a Rényi-like integral functional of hyperspherical harmonics, which can be expressed in terms of Rényi-like functionals of Gegenbauer polynomials $\mathcal{C}_{n-l-1}^{(\alpha'')}$ with $\alpha'' = D/2 +l- 1/2$ ; later on, we evaluate these Gegenbauer functionals at large $D$ by means of Theorem \ref{T3bis} (see Appendix \ref{B}), with emphasis in the circular and $(ns)$ states which are characterized by the hyperquantum numbers ($n,l=n-1, \{\mu\} = \{n-1\}$) and ($n,l=0, \{\mu\} = \{0\}$), respectively.\\
 
 Operating similarly in momentum space we can determine the momentum Rényi entropy of the system. In this space both the radial and angular parts of the momentum wave functions of the hydrogenic states are controlled by Gegenbauer polynomials as follows from the previous section. Consequently, the two radial and angular contributions to the momentum Rényi entropy are expressed in terms of Rényi-like functionals of Gegenbauer polynomials. 
\subsection{Rényi entropy in position space}
Let us obtain the position Rényi entropy of the probability density $\rho_{n,l,\{\mu\}}(\vec{r})$ given by (\ref{eq:denspos}), which according to (\ref{eq:renentropa}) is defined as
\begin{equation}
\label{eq:renentrop}
R_{q}[\rho_{n,l,\{\mu\}}] =  \frac{1}{1-q}\log W_{q}[\rho_{n,l,\{\mu\}}]; \quad 0<q<\infty, \,\, q \neq 1,
\end{equation}
where the symbol $W_{q}[\rho_{n,l,\{\mu\}}]$ denotes the entropic moments of the density 
\begin{eqnarray}
\label{eq:entropmom2}
W_{q}[\rho_{n,l,\{\mu\}}] &=& \int_{\mathbb{R}^D} [\rho_{n,l,\{\mu\}}(\vec{r})]^{q}\, d\vec{r}\nonumber\\ &=& \int\limits_{0}^{\infty}[\rho_{n,l}(\tilde{r})]^{q}\,r^{D-1}\,dr\times \Lambda_{l,\{\mu\}}(\Omega_{D-1}),
\end{eqnarray}
with
the angular part given by
\begin{equation}
\label{eq:angpart}
\Lambda_{l,\{\mu\}}(\Omega_{D-1}) = \int_{S^{D-1}}|\mathcal{Y}_{l,\{\mu\}}(\Omega_{D-1})|^{2q}\, d\Omega_{D-1}.
\end{equation}

Then, from Eqs. (\ref{eq:entropmom2}) and (\ref{eq:renentrop}) we can obtain the Rényi entropies of the $D$-dimensional hydrogenic state $(n,l,\{\mu\})$ as follows
\begin{equation}
\label{eq:renyihyd1}
R_{q}[\rho_{n,l,\{\mu\}}] = R_{q}[\rho_{n,l}]+R_{q}[\mathcal{Y}_{l,\{\mu\}}],
\end{equation}
where $R_{q}[\rho_{n,l}]$ denotes the radial part
\begin{equation}
\label{eq:renyihyd2}
R_{q}[\rho_{n,l}] = \frac{1}{1-q}\log \int_{0}^{\infty} [\rho_{n,l}(\tilde{r})]^{q} r^{D-1}\, dr,
\end{equation}
and  $R_{q}[\mathcal{Y}_{l,\{\mu\}}]$ denotes the angular part
\begin{equation}
\label{eq:renyihyd3}
R_{q}[\mathcal{Y}_{l,\{\mu\}}] = \frac{1}{1-q}\log \Lambda_{l,\{\mu\}}(\Omega_{D-1}).
\end{equation}
Here our aim is to determine the large-$D$ behavior of the Rényi entropy $R_{q}[\rho_{n,l,\{\mu\}}]$ when  all the hyperquantum numbers are fixed. According to (\ref{eq:renyihyd1}) this issue requires the knowledge at $D>>1$ of the radial and angular Rényi entropies, i.e. $R_{q}[\rho_{n,l}]$ and  $R_{q}[\mathcal{Y}_{l,\{\mu\}}]$ respectively, whose determination is done in the following.

\subsubsection{Radial position Rényi entropy }
\label{pos_rad_ren}

According to Eq. (\ref{eq:renyihyd2}), the radial Rényi entropy can be expressed as 
\begin{equation}
\label{eq:renyihyd4}
R_{q}[\rho_{n,l}] = \frac{1}{1-q}\log\left[\frac{\eta^{D(1-q)-q}}{2^{D(1-q)+q}Z^{D(1-q)}}N_{n,l}(D,q) \right],
\end{equation}
where $N_{n,l}(D,q)$ denotes the following $\mathfrak{L}_{q}$-norm of the Laguerre polynomials
\begin{equation}\label{eq:c1.21}
N_{n,l}(D,q)=\int\limits_{0}^{\infty}\left(\left[\widehat{\mathcal{L}}_{n-l-1}^{(\alpha)}(x)\right]^{2}\,w_{\alpha}(x)\right)^{q}\,x^{\beta}\,dx,
\end{equation}
with $\tilde{r} \equiv x$ and
\begin{equation}
\label{eq:condition}
\alpha=D +2l-2\,,\;l=0,1,2,\ldots,n-1,\, q>0\,\, \text{and}\,\, \beta=(2-D)q+D-1.
\end{equation}
We note that (\ref{eq:condition}) guarantees the convergence of
integral (\ref{eq:c1.21}); i.e., the condition $\beta+q\alpha= 2lq+D-1 > -1$ is always satisfied for physically meaningful values of the parameters. Moreover, the norm $N_{n,l}(D,q)$ can be rewritten as
\begin{eqnarray}
\label{eq:ren1}
N_{n,l}(D,q) &=& \int\limits_{0}^{\infty}\left(\left[\widehat{\mathcal{L}}_{n-l-1}^{(D +2l-2)}(x)\right]^{2}\,w_{D+2l-2}(x)\right)^{q}\,x^{2q-1+(1-q)D}\,dx\nonumber\\
&=& \left[\frac{\Gamma(n-l)}{\Gamma(n+l+D-2)}\right]^{q}\int\limits_{0}^{\infty}x^{D+2lq-1}e^{-qx} \left[\mathcal{L}_{n-l-1}^{(D +2l-2)}(x)\right]^{2q}\,dx.\nonumber \\
\end{eqnarray}

Then, the determination of the large-$D$ behavior of the radial Rényi entropy $R_{q}[\rho_{n,l}]$ requires the calculation of the asymptotics of the Laguerre functional $N_{n,l}(D,q)$ defined by (\ref{eq:c1.21}); that is, the evaluation of the Rényi-like integral functional given by (\ref{eq:ren1}) when $D >> 1$. We do it by applying Theorem \ref{T1bis} (see Appendix \ref{A}) to the functional $N_{n,l}(D,q)$ given by (\ref{eq:ren1}) with ($n,l$) fixed, obtaining for every non-negative $q \neq 1$ that

\begin{eqnarray}
\label{eq:ren6}
N_{n,l}(D,q) &\sim &  \left[\frac{\Gamma(n-l)}{\Gamma(n+l+D-2)}\right]^{q} \frac{\sqrt{2 \pi }}{[\Gamma(n-l)]^{2 q}} \left| q-1\right| ^{2 (n-l-1) q}\nonumber\\
& &\times e^{-D-2 l+2} (D+2 (l-1))^{D+2 q (n-1)-\frac{1}{2}} q^{-D-2 q (n-1)}\nonumber  \\
&=& \frac{\sqrt{2\pi} |q-1|^{2(n-l-1)q}}{\Gamma(n-l)^q}e^{-D-2l+2}\frac{(D+2l-2)^{D+2q(n-1)-\frac12}q^{-D-2q(n-1)}}{\Gamma(D+n+l-2)^q}\nonumber\\
&\sim& \frac{(2\pi)^{\frac{1-q}2} |q-1|^{2(n-l-1)q}}{\Gamma(n-l)^q}q^{-2q(n-1)}\left(\frac De\right)^{D(1-q)}q^{-D}D^{q(n-l+\frac12)-\frac12}\nonumber\\
\end{eqnarray}
where we have used the  Stirling's formula \cite{nist1} for the gamma function $\Gamma(x) = e^{-x}x^{x-\frac{1}{2}}(2\pi)^{\frac{1}{2}} \left[1+ \mathcal{O}\left(x^{-1}\right)\right]$.\\
 Then, Eqs. (\ref{eq:renyihyd4})-(\ref{eq:ren6}) allow us to find the following large-$D$ behavior for the radial R\'enyi entropy:
 \begin{eqnarray}
   \label{eq:ren8}
    R_{q}[\rho_{n,l}]
      & \sim &\frac{1}{1-q}\log \Bigg\{ \frac{(2\pi)^{\frac{1-q}2} |q-1|^{2(n-l-1)q}}{\Gamma(n-l)^q}q^{-2q(n-1)}\frac{ D^{D(1-q)-q}e^{(2n-3)(1-q)}}{4^{D(1-q)}Z^{D(1-q)}}\left(\frac De\right)^{D(1-q)}q^{-D}D^{q(n-l+\frac12)-\frac12} \Bigg\}\nonumber
    \\\nonumber
       & =&\frac{1}{1-q}\log \Bigg\{ \frac{(2\pi)^{\frac{1-q}2} |q-1|^{2(n-l-1)q}}{\Gamma(n-l)^q}\frac{e^{(2n-3)(1-q)}}{q^{2q(n-1)}}\left(\frac {D^2}{4Ze}\right)^{D(1-q)}q^{-D}D^{q(n-l-\frac12)-\frac12} \Bigg\}\nonumber
       \\
   \end{eqnarray}
 which can be rewritten as 
  \begin{equation}
 \label{eq:ren9}
R_q[\rho_{n,l}]\sim 2D \log \left[D\right]+D \log \left[\frac{ q^{\frac1{q-1}}}{4Ze}\right]+\frac{q(n-l-\frac12)-\frac12}{1-q}\log D+\frac1{1-q}\log \mathcal{F}(n,l,q),
  \end{equation}
 where $\mathcal{F}(n,l,q) = \frac{(2\pi)^{\frac{1-q}2} |q-1|^{2(n-l-1)q}}{\Gamma(n-l)^q}\frac{e^{(2n-3)(1-q)}}{q^{2q(n-1)}}$. Further terms in this asymptotic expansion can be obtained by means of Theorem \ref{T1bis} (see Appendix \ref{A}).
 \\
 
 Note that, since $q^\frac1{q-1}\to e$ when $q \to 1$, we have the following conjecture for the value of the Shannon entropy  
   \begin{equation}
   S[\rho_{n,l}]\sim 2D \log \left[D\right]-D \log \left[4Z\right],
   \end{equation} 
which can be numerically shown to be correct. However a more rigorous expression for this quantity remains to be proved.\\
 
Then, according to Eq. (\ref{eq:renyihyd1}), to fix the total Rényi entropy $R_{q}[\rho_{n,l,\{\mu\}}]$ at large $D$ it only remains the evaluation of the corresponding large-$D$ behavior of the angular part $R_{q}[\mathcal{Y}_{l,\{\mu\}}]$ which will be done in the following.

\subsubsection{Angular Rényi entropy}
\label{pos_ang_ren}

Here we will effort to calculate the large-$D$ behavior of the angular part $R_{q}[\mathcal{Y}_{l,\{\mu\}}]$ of the total position and momentum Rényi entropies defined by Eq. (\ref{eq:renyihyd3}).
Therein, according to (\ref{eq:hyperspherarm}) and (\ref{eq:angpart}), the Rényi-like functional $\Lambda_{l,\{\mu\}}(\Omega_{D-1})$ of the hypersherical harmonics can be expressed as
%
\begin{eqnarray}
\label{eq:reha1}
\Lambda_{l,\{\mu\}}(\Omega_{D-1}) &=& \int_{S^{D-1}}|\mathcal{Y}_{l,\{\mu\}}(\Omega_{D-1})|^{2q}\, d\Omega_{D-1}\nonumber\\
&=& \mathcal{N}_{l,\{\mu \}}^{2q} \int_{S^{D-1}} \prod_{j=1}^{D-2}[C^{(\alpha_{j}+\mu_{j+1})}_{\mu_{j}-\mu_{j+1}}(\cos \theta_{j})]^{2q}(\sin\theta_{j})^{2q\mu_{j+1}}\, d\Omega_{D-1}\nonumber \\
&=& 2\pi\mathcal{N}_{l,\{\mu \}}^{2q}\prod_{j=1}^{D-2}\int_{0}^{\pi}[C^{(\alpha_{j}+\mu_{j+1})}_{\mu_{j}-\mu_{j+1}}(\cos \theta_{j})]^{2q}(\sin\theta_{j})^{2q\mu_{j+1}+2\alpha_j}\, d\theta_j\nonumber
\end{eqnarray}
where the normalization constant $\mathcal{N}_{l,\{\mu \}}$ is given by (\ref{eq:normhypersphar}). Moreover, note that
the integrals within the product are Rényi-like functionals of Gegenbauer polynomials of the type considered in Theorem \ref{T3bis} (see Appendix \ref{B}). \\
To calculate the dominant term of $\Lambda_{l,\{\mu\}}(\Omega_{D-1})$ at large $D$ we use the Theorem \ref{T3bis} at zeroth-order or, what is equivalent, 
we use the following limiting expressions of the Gegenbauer polynomials to monomials (see \cite{nist1}, Eq. $18.6.4$)
\begin{equation}
\label{eq:renp4}
\lim_{\alpha' \to \infty} (2\alpha')^{-k}\mathcal{C}_{k}^{(\alpha')}(x) = \frac{x^{k}}{k!},
\end{equation}
which allows us to find 
\begin{eqnarray}
\label{eq:reha2}
\Lambda_{l,\{\mu\}}(\Omega_{D-1}) &\sim& 2\pi \mathcal{N}_{l,\{\mu \}}^{2q}\prod_{j=1}^{D-2} \frac{[2(\alpha_{j}+\mu_{j+1})]^{2q(\mu_{j}-\mu_{j+1})}}{[\Gamma(\mu_{j}-\mu_{j+1}+1)]^{2q}}\int_{0}^{\pi} (\cos\theta_{j})^{2q(\mu_{j}-\mu_{j+1})}(\sin\theta_{j})^{2(q\mu_{j+1}+\alpha_{j})}\, d\theta_{j}\nonumber \\
&=& 2\pi \,\mathcal{N}_{l,\{\mu \}}^{2q}\prod_{j=1}^{D-2} \frac{[2(\alpha_{j}+\mu_{j+1})]^{2q(\mu_{j}-\mu_{j+1})}}{[\Gamma(\mu_{j}-\mu_{j+1}+1)]^{2q}}B\left(q\mu_{j+1}+\frac{1}{2}+\alpha_{j}, q(\mu_{j}-\mu_{j+1})+\frac{1}{2} \right)\nonumber \\\nonumber
&= & (2\pi)^{1+(1-D)q}\left(\prod_{j=1}^{D-2}\frac{\Gamma\big(q(\mu_j-\mu_{j+1})+\frac12\big)}{\Gamma\big(\mu_j-\mu_{j+1}+1\big)^q}\right)\left(\prod_{j=1}^{D-2}(\alpha_j+\mu_j) (\alpha_j+\mu_{j+1})^{2(\mu_j-\mu_{j+1})}\right)^q\\\nonumber
&&\times
\left(\prod_{j=1}^{D-2}\frac{\Gamma(\alpha_j+q\mu_{j+1}+\frac12)}{\Gamma(\alpha_j+q\mu_j+1)}\right)\left(\prod_{j=1}^{D-2}4^{q(\alpha_j+\mu_j)}\frac{\Gamma(\alpha_j+\mu_{j+1})^{2q}}{\Gamma(2\alpha_j+\mu_{j+1}+\mu_j)^q}\right).\\
\end{eqnarray}
Further algebraic manipulations, which are detailed in Appendix \ref{details}, have allowed us to obtain that
\begin{eqnarray}
\label{eq:rehabis}
\Lambda_{l,\{\mu\}}(\Omega_{D-1}) &\sim& \left( 2^{1-q}\frac{\Gamma(1+q\mu_{D-1})}{\Gamma(1+\mu_{D-1})^q} \tilde{\mathcal M}(D,q,\{\mu\})\right)\\\nonumber
&&\times\left(\tilde{\mathcal E}(D,\{\mu\})^q \pi^{\frac D2(1-q)}\frac{\Gamma(\frac D2+l)^q}{\Gamma(\frac D2+q\,l)}\right)
\end{eqnarray}
where 
\begin{equation}\label{eq:Mtilde}
\tilde{\mathcal M}(D,q,\{\mu\})\equiv 4^{q(l-\mu_{D-1})} \pi^{1-\frac D2}\prod_{j=1}^{D-2}\frac{\Gamma\big(q(\mu_j-\mu_{j+1})+\frac12\big)}{\Gamma\big(\mu_j-\mu_{j+1}+1\big)^q}
\end{equation}
and 
\begin{equation}\label{eq:Etilde}
\tilde{\mathcal E}(D,\{\mu\})\equiv\prod_{j=1}^{D-2} \frac{(\alpha_j+\mu_{j+1})^{2(\mu_j-\mu_{j+1})}}{(2\alpha_j+2\mu_{j+1})_{\mu_j-\mu_{j+1}}}\frac{1}{(\alpha_j+\mu_{j+1})_{\mu_j-\mu_{j+1}}}.
\end{equation}
Note that $\tilde{\mathcal E}=\tilde{\mathcal M}=1$ for any configuration with $\mu_1=\mu_2=\cdots=\mu_{D-1}$. 
Finally, as explained in Appendix \ref{details}, we have the following expression 
\begin{eqnarray}\label{eq:Rq_ang}
R_{q}[\mathcal{Y}_{l,\{\mu\}}] &\sim & \frac1{1-q}\log\left(\frac{\Gamma\left(\frac D2+l\right)^q}{\Gamma\left(\frac D2+ql\right)}\right) +\frac D2\log\pi\nonumber\\ 
&&+\frac1{1-q}\log\left(\tilde{\mathcal E}(D,\{\mu\})^q\tilde{\mathcal M}(D,q,\{\mu\})\frac{\Gamma(1+q\mu_{D-1})}{\Gamma(1+\mu_{D-1})^q}2^{1-q}\right) \nonumber\\
&\sim&-\log\left(\Gamma\left(\frac D2\right)\right)+\frac D2\log\pi+\frac 1{1-q}\log\left(\tilde{\mathcal E}(D,\{\mu\})^q\tilde{\mathcal M}(D,q,\{\mu\})\frac{\Gamma(1+q\mu_{D-1})}{\Gamma(1+\mu_{D-1})^q} 2^{1-q}\right)\nonumber\\
&\sim&-\frac D2 \log D+\frac D2\log(2\pi e)+\frac12\log D+\frac 1{1-q}\log\left(\tilde{\mathcal E}(D,\{\mu\})^q\tilde{\mathcal M}(D,q,\{\mu\})\frac{\Gamma(1+q\mu_{D-1})}{\Gamma(1+\mu_{D-1})^q}\left(\frac\pi2\right)^{\frac{q-1}2}\right)\nonumber\\
\end{eqnarray}
for the angular R\'enyi entropy of the generic hydrogenic state with hyperquantum numbers $(l,\{\mu\})$, which holds for every non-negative $q \neq 1$.\\
For completeness, we will determine this asymptotic behavior in a more complete manner for some physically-relevant and experimentally accessible states like the $(ns)$ and circular ones, which are described by the hyperquantum numbers  ($n,l=0, \{\mu\} = \{0\}$) and ($n, l=n-1$, $\{\mu\} = \{n-1\}$), respectively. First we obtain from (\ref{eq:reha2}) the following values 
\begin{eqnarray}
\label{eq:reha3}
\Lambda_{0,\{0\}}(\Omega_{D-1}) &\sim&2\pi\mathcal N_{0,\{0\}}^{2q}\prod_{j=1}^{D-2}\frac{\Gamma(\frac12)\Gamma(\alpha_j+\frac12)}{\Gamma(\alpha_j+1)}=\left(\frac{2\pi^{\frac{D}{2}}}{\Gamma\left(\frac D2\right)}\right)^{1-q}
	\end{eqnarray}
	

and 
\begin{eqnarray}
\label{eq:reha4}
\Lambda_{n-1,\{n-1\}}(\Omega_{D-1}) &\sim& 2\pi\mathcal N_{n-,1\{n-1\}}^{2q}\prod_{j=1}^{D-2}\frac{\Gamma(\frac12)\Gamma(\alpha_j+q(n-1)\frac12)}{\Gamma(\alpha_j+q(n-1)+1)}\nonumber\\
&=&\left(2\pi^\frac D2\right)^{1-q}\prod_{j=1}^{D-2}\left(\frac{\Gamma(\alpha_j+n)}{\Gamma(\alpha_j+n-\frac12)}\right)^q\frac{\Gamma(\alpha_j+q(n-1)+\frac12)}{\Gamma(\alpha_j+q(n-1)+1)}\nonumber\\
&=& \left(\frac1{2\pi^\frac D2}\right)^{q-1}\frac{\left(\left(n\right)_{\frac D2-1}\right)^{q}}{\left(1+q(n-1)\right)_{\frac D2-1}},\\\nonumber
\end{eqnarray}
respectively. Note that $(x)_{a}=\frac{\Gamma(x+a)}{\Gamma(x)}$ is the well-known Pochhammer symbol. Then, from Eqs. (\ref{eq:renyihyd3}), (\ref{eq:reha3}) and (\ref{eq:reha4}), we have that the angular part of the Rényi entropy at large $D$ is given by
\begin{eqnarray}
\label{eq:reha5}
R_{q}[\mathcal{Y}_{0,\{0\}}]\sim \log \left(\frac{2\pi^\frac D2}{\Gamma\left(\frac D2\right)}\right)
\end{eqnarray}
and  
\begin{eqnarray}
\label{eq:reha6}
R_{q}[\mathcal{Y}_{n-1,\{n-1\}}]  &\sim& \frac 1{1-q}\log\left(\left(\frac1{2\pi^\frac D2}\right)^{q-1}\frac{\left(\left(n\right)_{\frac D2-1}\right)^{q}}{\left(1+q(n-1)\right)_{\frac D2-1}}\right)
\end{eqnarray}
for the $(ns)$ and circular states, respectively. Note that for very large $D$ the dominant term of the angular R\'enyi entropy of these two classes of physical states is the same; namely, $-\log\left(\Gamma\left(\frac D2\right)\right)+\frac D2\log\pi$. Moreover and most interesting: this is true for any hydrogenic state by taking into account the general expression (\ref{eq:Rq_ang}). This observation allows us to conjecture the expression
 \begin{equation}
 S[\mathcal{Y}_{l,\{\mu\}}]\sim-\log\left(\Gamma\left(\frac D2\right)\right)+\frac D2\log{\pi}.
  \end{equation} 
for the large-$D$ behavior of the angular Shannon entropy of the hydrogenic states.
  
\subsubsection{Total position Rényi entropy}

To obtain the total Rényi entropy $R_q[\rho_{n,l,\{\mu\}}]$ in position space for a general $(n,l,\{\mu\})$-state, according to \eqref{eq:renyihyd1}, we have to sum up the radial and angular contributions given by \eqref{eq:ren9} and (\ref{eq:Rq_ang}), respectively. Then, we obtain that
\begin{eqnarray}\label{eq:TotSpa}\nonumber
 R_q[\rho_{n,l,\{\mu\}}]&\sim& \log\left(\frac{D^{2D}}{\Gamma\left(\frac D2\right)}\right)+D\log\left(\frac{q^{\frac1{q-1}}\sqrt{\pi}}{4Ze}\right)+\frac{q(n-l- \frac12)-\frac12}{1-q}\log D\\\nonumber
 &&+\frac 1{1-q}\log\left(\tilde{\mathcal E}(D,\{\mu\})^q\tilde{\mathcal M}(D,q,\{\mu\})\mathcal F(n,l,q)\,\frac{\Gamma(1+q\mu_{D-1})}{\Gamma(1+\mu_{D-1})^q}\,\left(\frac\pi2\right)^{\frac{q-1}2}\right) \\\nonumber 
 &\sim& \frac32D\log D+D\log\left(\frac{q^{\frac1{q-1}}}{Z}\sqrt{\frac\pi{8e}}\right)+\frac{q(n-l-1)}{1-q}\log D\\\nonumber
  &&+\frac 1{1-q}\log\left(\tilde{\mathcal E}(D,\{\mu\})^q\tilde{\mathcal M}(D,q,\{\mu\})\mathcal F(n,l,q)\,\frac{\Gamma(1+q\mu_{D-1})}{\Gamma(1+\mu_{D-1})^q}\,\left(\frac\pi2\right)^{\frac{q-1}2}\right)\\
\end{eqnarray}
which holds for every non-negative $q\neq 1$. Now, for completeness we calculate this quantity in an explicit manner for the $(ns)$ and circular states, which both of them include the ground state. For the $(ns)$ and circular states we have the asymptotical expressions
\begin{eqnarray}\label{eq:TotSpaCir}\nonumber
 R_q[\rho_{n,0,\{0\}}]&\sim&\log\left(\frac{D^{2D}}{\Gamma\left(\frac D2\right)}\right)+D\log\left(\frac{q^{\frac1{q-1}}\sqrt{\pi}}{4Ze}\right)+\frac{q(n-1)}{1-q}\log D+\frac 1{1-q}\log\left(\mathcal F(n,0,q)\right)-\frac12\log \frac \pi2, \\
\end{eqnarray}
(with $\mathcal{F}(n,0,q) = \frac{\left| q-1\right| ^{2 (n-1) q}}{(2 \pi )^{\frac{1}{2}(q-1)}[(n-1)!]^q }$) and 
\begin{eqnarray}\label{eq:TotSpans}\nonumber
 R_q[\rho_{n,n-1,\{n-1\}}]&\sim& \frac1{1-q}\log\left(\frac{\left(\left(n\right)_{\frac D2-1}\right)^{q}}{\left(1+q(n-1)\right)_{\frac D2-1}}\right)+2D\log D\\\nonumber
 &&+ D\log\left(\frac{q^{\frac1{q-1}}\sqrt{\pi}}{4Ze}\right)+\frac 1{1-q}\log\left(\mathcal F(n,n-1,q)\,\frac{\Gamma(1+q(n-1))}{\Gamma(n)^q}\right)-\frac12\log \frac \pi2\\
 &\sim&\log\left(\frac{D^{2D}}{\Gamma\left(\frac D2\right)}\right)+D\log\left(\frac{q^{\frac1{q-1}}\sqrt{\pi}}{4Ze}\right)+\frac 1{1-q}\log\left(\frac{\Gamma(1+q(n-1))}{q^{2q(n-1)}\Gamma(n)^q}\right)+2n+\log2-3,\nonumber\\
\end{eqnarray}
respectively. Most interesting, we realize that the large-$D$ behavior of the total R\'enyi entropy in the position space for the hydrogenic ground state $R_q[\rho_{1,0,\{0\}}]$ is given by the last expression above indicated.\\

Finally, from \eqref{eq:TotSpa} one can conjecture that the Shannon entropy $S[\rho_{n,l,\{\mu\}}]$ in position space for a general $(n,l,\{\mu\})$-state is given by
\begin{eqnarray}\label{eq:Cpshannon}
 S[\rho_{n,l,\{\mu\}}]&\sim& \log\left(\frac{D^{2D}}{\Gamma\left(\frac D2\right)}\right)+D\log\left(\frac{\sqrt{\pi}}{4Z}\right)\\\nonumber
 &\sim& \frac32D\log D+D\log\left(\frac{\sqrt{e\pi}}{\sqrt 8Z}\right)
\end{eqnarray}
but this is somehow risky because of the unknown ($q\to1$)-behavior of the angular part, coming from the second line of Eq. (\ref{eq:TotSpa}).
\subsection{Rényi entropy in momentum space}

Let us now determine the momentum Rényi entropy of the probability density $\gamma_{n,l,\{\mu\}}(\vec{p})$ given by (\ref{eq:momdens}), which is defined as
\begin{equation}
\label{eq:momrenentrop}
R_{q}[\gamma_{n,l,\{\mu\}}] =  \frac{1}{1-q}\log W_{q}[\gamma_{n,l,\{\mu\}}]; \quad 0<q<\infty,\,\, q \neq 1,
\end{equation}
where
\begin{equation}
\label{eq:entropmom3}
W_{q}[\gamma_{n,l,\{\mu\}}] = \int_{\mathbb{R}^D} [\gamma_{n,l,\{\mu\}}(\vec{p})]^{q}\, d\vec{p}	
\end{equation}
denote the momentum entropic moments. Then, operating in a similar way as in position space, we obtain that
\begin{equation}
\label{eq:renyihyd5}
R_{q}[\gamma_{n,l,\{\mu\}}] = R_{q}[\gamma_{n,l}]+R_{q}[\mathcal{Y}_{l,\{\mu\}}],
\end{equation}
where $R_{q}[\gamma_{n,l}]$ denotes the radial part
\begin{equation}
\label{eq:renyihyd6}
R_{q}[\gamma_{n,l}] = \frac{1}{1-q}\log \int_{0}^{\infty} [\mathcal{M}_{n,l}(p)]^{2q} p^{D-1}\, dp,
\end{equation}
and $R_{q}[\mathcal{Y}_{l,\{\mu\}}]$ denotes the angular part given by (\ref{eq:renyihyd3}).\\

Since the angular entropy $R_{q}[\mathcal{Y}_{l,\{\mu\}}]$ has been already calculated and discussed, it only remains to determine at large $D$ the radial Rényi entropy $R_{q}[\gamma_{n,l}]$, given by (\ref{eq:renyihyd6}), being all the hyperquantum numbers fixed. 

\subsubsection{Radial momentum Rényi entropy}
\label{mom_rad_ren}

To find the radial momentum Rényi entropy (\ref{eq:renyihyd6}) at $D>> 1$, we first rewrite for convenience the radial part of the wave function, $\mathcal{M}_{n,l}(p)$, given by (\ref{eq:momdens}) as
\begin{eqnarray}
\label{eq:renp2}
\mathcal{M}_{n,l}(p) = \left(\frac{\eta}{Z}\right)^{\frac{D}{2}}(1+y)^{\frac{3}{2}}\left(\frac{1+y}{1-y}\right)^{\frac{D-2}{4}}\sqrt{w_{\alpha}(y)}\,\widehat{\mathcal{C}}^{(\alpha)}_{n-l-1}(y)\\\nonumber
=\left(\frac{\eta}{Z}\right)^{\frac{D}{2}}A(n,l;D)^\frac12(1-y)^{\frac{l}{2}}(1+y)^{\frac{D+l+1}{2}}{\mathcal{C}}^{(l+\frac{D-1}2)}_{n-l-1}(y),
\end{eqnarray}
with $y=\frac{1-\eta^{2}\tilde{p}^{2}}{1+\eta^{2}\tilde{p}^{2}}$, $\tilde{p}=p/Z$, $\alpha = l+ \frac{D-1}{2}$, and $\widehat{\mathcal{C}}^{(\alpha)}_{n-l-1}(x)$ denotes the orthonormal Gegenbauer polynomials with respect to the weight function $w_{\alpha}(x) = (1-x^{2})^{\alpha-\frac{1}{2}}$, so that $\widehat{\mathcal{C}}^{(l+\frac{D-1}2)}_{n-l-1}(x)=A(n,l;D)^\frac12{\mathcal{C}}^{(l+\frac{D-1}2)}_{n-l-1}(x)$, where the constant
\begin{equation}
	A(n,l;D) = \frac{(n-l-1)!(n+\frac{D-3}{2})[\Gamma(l+\frac{D-1}{2})]^{2}}{2^{2-2l-D}\pi\Gamma(n+l+D-2)}.
\end{equation} 

The radial momentum Rényi entropy (\ref{eq:renyihyd6}) together with (\ref{eq:renp2}) can be expressed as
\begin{eqnarray}
\label{eq:renp1}
R_{q}[\mathcal{M}_{n,l}] &=& -D\log\frac{\eta}{Z}+\frac{1}{1-q}\left\{q\log A(n,l;D) +\log I_{n,l}(q,D)\right\}\nonumber \\
\end{eqnarray}

and the following Rényi-like functional of Gegenbauer polynomials
\begin{equation}
I_{n,l}(q,D) = \int_{-1}^{1} (1-y)^{lq-1+\frac{D}{2}}(1+y)^{(l+1)q-1+(q-\frac{1}{2})D}[\mathcal{C}^{(l+\frac{D-1}{2})}_{n-l-1}(y)]^{2q}\, dy.
\end{equation}
It only remains to find the large-$D$ behavior of the two terms in (\ref{eq:renp1}) when ($n,l,q$) are fixed.
The asymptotic estimate of $A(n,l;D) $ turns out to be given by
\begin{equation}
\label{eq:asympAnl}
	A(n,l;D) \sim \frac{\Gamma (n-l)}{\sqrt{2 \pi }}D^{l-n+\frac{3}{2}} .
\end{equation} 
On the other hand, the behavior of $I_{n,l}(q,D)$ at large $D$ can be obtained from Theorem \ref{T3bis} (see Appendix \ref{B}) by studying the large-$\alpha$ behavior of the integral $J_2(a,b,c,d,\kappa,m';\alpha)$ with the parameters $a=l(q-1)-\frac12, \, b= l (1-q)+2q-\frac32,\, c=1,\, d=2q-1,\,\,\kappa = 2q,\,m' =n-l-1$ and $\alpha = l+\frac{D-1}{2}$. Note that the condition $c<d$ of the theorem provokes that $q>1$.
%
We have found that
\begin{eqnarray}
\label{eq:asyInl}
I_{n,l}(q,D)  \sim\left(\frac{(2q-1)^{2q-1}}{q^{2q}}\right)^{\frac D2}\left(\frac{\Gamma(\frac D2+n-\frac32)}{\Gamma(\frac D2+l-\frac12)}\right)^{2q}(D+2l-1)^{-\frac12}Q_0(n,l,q)
\nonumber\\
\sim \left(\frac{(2q-1)^{2q-1}}{q^{2q}}\right)^{\frac D2}\left(\frac D2\right)^{2q(n-l-1)}(D+2l-1)^{-\frac12}Q_0(n,l,q)
\nonumber\\
\sim \left(\frac{(2q-1)^{2q-1}}{q^{2q}}\right)^{\frac D2} D^{2q(n-l-1)-\frac12}\frac{Q_0(n,l,q)}{4^{q(n-l-1)}}  
\end{eqnarray}
for $q>1$ and where  $Q_{0}(q)$ is given by

\begin{equation}
\label{eq:Q0}
Q_{0}(q,n,l) =\frac{\sqrt{2\pi}\,4^{q(n-l-1)}}{\Gamma(n-l)^{2q}}\frac{(2 q-1)^{q(l+1)-\frac12}(q-1)^{2 q (n-l-1)}}{ q^{ q(2n-1)-\frac{1}{2}}}.
\end{equation}
 Using (\ref{eq:asympAnl}) and (\ref{eq:asyInl}) we obtain the following behavior at large $D$ for the radial part of the momentum Rényi entropy:

\begin{eqnarray}
\label{eq:renp7}\nonumber
R_{q}[\mathcal{M}_{n,l}] &\sim& 
-D\log\left(\frac \eta{Z}\right)+\frac D{1-q}\log \sqrt\frac{(2q-1)^{2q-1}}{q^{2q}}+\frac{q(n-l-\frac12)-\frac12}{1-q}\log D\\ \nonumber
&&+\frac1{1-q}\log \overline{Q}_0(q,n,l)\\
&\sim& 
-D\log\left(\frac D{2Z}\right)+\frac D{1-q}\log \sqrt\frac{(2q-1)^{2q-1}}{q^{2q}}+\frac{q(n-l-\frac12)-\frac12}{1-q}\log D\\ \nonumber
&&+\frac1{1-q}\log \overline{Q}_0(q,n,l)
\end{eqnarray}
with $\overline{Q}_0(q,n,l)=\frac{\Gamma(n-l)^q}{(2\pi)^\frac q2\,4^{q(n-l-1)}}Q_0(q,n,l)=\frac{{(2\pi)^{\frac{1-q}2}}}{\Gamma(n-l)^{q}}\frac{(2 q-1)^{q(l+1)-\frac12}(q-1)^{2 q (n-l-1)}}{ q^{ q(2n-1)-\frac{1}{2}}} $ 
(where we have used in the second expression that $\eta=n+\frac{D-3}{2} \sim \frac{D}{2}$ for fixed $n$) for a general $(n,l)$-state. Note that in the limit $q \to 1$ this expression suggests that the behavior of the radial Shannon entropy in the momentum space can be conjectured at large $D$ as
\begin{equation}
S[\mathcal{M}_{n,l}]\sim -D\log\left(\frac D{2Z}\right).	
\end{equation}
Then, according to Eq. (\ref{eq:renyihyd5}), to fix the large $D$-behavior of the total Rényi entropy $R_{p}[\gamma_{n,l,\{\mu\}}]$ it only remains the evaluation of the corresponding behavior of the angular part $R_{p}[\mathcal{Y}_{l,\{\mu\}}]$ which was found in \eqref{eq:Rq_ang}.

\subsubsection{Total momentum Rényi entropy}

To obtain the total Rényi entropy in momentum space for a general $(n,l,\{\mu\})$-state of a large-dimensional hydrogenic system we have to sum up the radial and angular contributions, given by \eqref{eq:renyihyd5}, and then to take into account the final expressions \eqref{eq:renp7} and \eqref{eq:Rq_ang} for these contributions. Then, it follows the expression
\begin{eqnarray}\label{eq:TMrenyi}\nonumber
R_q[\gamma_{n,l,\{\mu\}}]&\sim&-\log\left(\frac{\eta^D\,\Gamma\left(\frac D2\right)}{Z^D}\right) + D\log\left(\sqrt{\pi}\left(\frac{(2q-1)^{2q-1}}{q^{2q}}\right)^{\frac1{2-2q}}\right)\\
&&+\frac{{q(n-l- \frac12)-\frac12}}{1-q}\log D +\frac1{1-q}\log\left(\tilde{\mathcal E}(D,\{\mu\})^q\tilde{\mathcal M}(D,q,\{\mu\})\overline{Q}_0(q,n,l) \pi^\frac{q-1}2\right)\nonumber\\
&\sim&-\frac 32 D\log D + D\log\left(Z\sqrt{8e\pi}\left(\frac{(2q-1)^{2q-1}}{q^{2q}}\right)^{\frac1{2-2q}}\right)\nonumber\\
	&&+\frac{{q(n-l-1)}}{1-q}\log D +\frac1{1-q}\log\left(\tilde{\mathcal E}(D,\{\mu\})^q\tilde{\mathcal M}(D,q,\{\mu\})\overline{Q}_0(q,n,l) \pi^\frac{q-1}2\right),\nonumber\\
\end{eqnarray}
(with $q\neq 1$ and $\eta \sim \frac{D}{2}$) for the large-$D$ behavior of the total momentum Rényi entropy of the generic hydrogenic state $(n,l,\{\mu\})$, where the symbols $\tilde{\mathcal M}(D,q,\{\mu\})$ and $\tilde{\mathcal E}(D,\{\mu\})$ are defined in Eqs. \eqref{eq:Mtilde} and \eqref{eq:Etilde}, respectively . For completeness and illustration, let us give  in a more complete manner the behavior of this quantity at $D>>1$ for some particular quantum states such as the $(ns)$ and circular states. For the $(ns)$-states we found

\begin{eqnarray}\label{eq:TotMomns}\nonumber
 R_q[\gamma_{n,0,\{0\}}]&\sim&-\frac 32 D\log D + D\log\left(Z\sqrt{8e\pi}\left(\frac{(2q-1)^{2q-1}}{q^{2q}}\right)^{\frac1{2-2q}}\right)\\
&&+\frac{{q(n-1)}}{1-q}\log D +\frac1{1-q}\log\left(\overline{Q}_0(q,n,0) \pi^\frac{q-1}2\right)\nonumber\\
\end{eqnarray}
with
\begin{equation}
\overline{Q}_0(q,n,0)=\frac{{(2\pi)^{\frac{1-q}2}}}{\Gamma(n)^{q}}\frac{(2 q-1)^{q-\frac12}(q-1)^{2 q (n-1)}}{ q^{ q(2n-1)-\frac{1}{2}}}.
\end{equation}
And for the circular states we obtained the following large-$D$ behavior
\begin{eqnarray}\label{eq:TotMomCir}\nonumber
R_q[\gamma_{n,n-1,\{n-1\}}]
&\sim&-\frac 32 D\log D + D\log\left(Z\sqrt{8e\pi}\left(\frac{(2q-1)^{2q-1}}{q^{2q}}\right)^{\frac1{2-2q}}\right)\\
&&+\frac1{1-q}\log\left(\overline{Q}_0(q,n,n-1) \pi^\frac{q-1}2\right)\nonumber\\
\end{eqnarray}
with 
\begin{equation}
\overline{Q}_0(q,n,n-1)=(2\pi)^{\frac{1-q}2}\frac{(2 q-1)^{qn-\frac12}}{ q^{ q(2n-1)-\frac{1}{2}}}.
\end{equation}
Note now that either from \eqref{eq:TotMomns} or from \eqref{eq:TotMomCir} with $n=1$ we have the following large-$D$ behavior 
\begin{eqnarray}\label{eq:TotMomFun}\nonumber
R_q[\gamma_{1,0,\{0\}}] &\sim& -\frac 32 D\log D + D\log\left(Z\sqrt{8e\pi}\left(\frac{(2q-1)^{2q-1}}{q^{2q}}\right)^{\frac1{2-2q}}\right)\\
&&+\frac1{1-q}\log\left(\overline{Q}_0(q,1,0) \pi^\frac{q-1}2\right)\nonumber\\
\end{eqnarray}
with $q\neq 1$ and $\overline{Q}_0(q,1,0)=(2\pi)^{\frac{1-q}2}\left(2-\frac1q\right)^{q-\frac12}$
for the total momentum R\'enyi entropy of the ground hydrogenic state.

Finally, from \eqref{eq:TMrenyi} one can conjecture that the Shannon entropy $S[\rho_{n,l,\{\mu\}}]$ in momentum space for a general $(n,l,\{\mu\})$-state is given by
\begin{equation}\label{eq:ShannonMom}
S[\gamma_{n,l,\{\mu\}}]\sim -\frac 32 D\log D + D\log\left(Z\sqrt{8e\pi}\right)
\end{equation}
in the limiting case $q\rightarrow 1$.

\subsection{Position-momentum R\'enyi-entropy-based uncertainty sum}
From Eqs. \eqref{eq:TotSpa} and \eqref{eq:TMrenyi} we can obtain the dominant term for the joint position-momentum R\'enyi-entropy-based uncertainty sum of a large-dimensional hydrogenic system for a pair of parameters $p$ and $q$ which fulfill the Holder conjugacy relation $\frac1p+\frac1q=2$. We found that
\begin{eqnarray}
R_q[\rho_{n,l,\{\mu\}}]+R_p[\gamma_{n,l,\{\mu\}}] &\sim& D\log\left[\pi\left(\frac{(2p-1)^{(2p-1)}}{p^{2p}}\right)^{\frac1{2-2p}} q^{\frac1{q-1}}\right]\nonumber\\
&= & D\log\left[2\pi\,(2p)^\frac1{2p-2}\,(2q)^\frac1{2q-2}\right],\quad q\neq1
\end{eqnarray}
for all $(n,l,\{\mu\})$-states, which saturates the known position-momentum R\'enyi-entropy-based uncertainty relation \cite{bialynicki2,vignat,zozor2008}. Note that out of the so called conjugacy curve (i.e., for arbitrary positive pairs of values of $p$ and $q$), there is a dependence at second order on the quantum numbers $n$ and $l$; this dependence just disappears onto the conjugation line. A more detailed study of this behaviour out of the conjugacy curve remain as an open problem. 
Finally, the conjectured expressions for Shannon entropy in both spaces (\ref{eq:Cpshannon}), (\ref{eq:ShannonMom})  allows one to write
\begin{eqnarray}
S[\rho_{n,l,\{\mu\}}]+S[\gamma_{n,l,\{\mu\}}] &\sim& D\log\left[2\pi e\right]
\end{eqnarray}
which saturates the general Bialynicki-Birula-Mycielski entropic relation \cite{bialynicki2,bialynicki3}. 

\section{Conclusions}
In this work we have determined the large-$D$ behavior of the position and momentum R\'enyi entropies of the $D$-dimensional hydrogenic states at large $D$ in terms of the state's hyperquantum numbers and the nuclear charge $Z$ of the system. We have used a recent constructive methodology which allows for the calculation of some Rényi-like integral functionals of Laguerre $\mathcal{L}^{(\alpha)}_{k}(x)$ and Gegenbauer $\mathcal{C}^{(\alpha'')}_{k}(x)$ polynomials with a fixed degree $k$ and large values of the parameters $\alpha$ and $\alpha''$.  This has been possible because the hydrogenic states are controlled by the Laguerre and Gegenbauer polynomials in position space, and by the Gegenbauer polynomials in momentum space, keeping in mind that the hyperspherical harmonics (which determine the angular part of the wave functions in the two conjugated spaces) can be expressed in terms of the latter polynomials. Then, simple expressions of these quantities for some specific classes of hydrogenic states ($ns$ and circular states), which include the ground state, are given. Moreover, as a byproduct, our results reach the saturation of the known position-momentum R\'enyi-entropy-based uncertainty relations. To this respect we should keep in mind that we are assuming that the dimensionality is very large and the hyperquantum numbers are small. The exceptional case when both dimensionality and hyperquantum numbers are simultaneously large has not yet been explored; in particular, we cannot assure saturation.

We should highlight that to find the Shannon entropies of the large-dimensional hydrogenic systems have not yet been possible with the present methodology, although the dominant term  has been conjectured. A rigorous proof remains open.

Finally, let us mention that it would be very relevant for many quantum-mechanical problems other than the hydrogenic ones (e.g., the harmonic systems) the determination of the behavior of integral functionals of Rényi and Shannon types for hypergeometric polynomials other than the Laguerre and Gegenbauer ones at large values of the polynomials' parameters and fixed degrees. This is yet another open problem for the future.

\section*{Acknowledgments}
This work has been partially supported by the Projects FQM-7276 and FQM-207 of the Junta de Andaluc\'ia and the MINECO (Ministerio de Economia y Competitividad)-FEDER (European Regional Development Fund) grants
 FIS2014- 54497P and FIS2014-59311-P. I. V. Toranzo acknowledges the support of ME under the program FPU. 
 N. M. Temme acknowledges financial support from 
{\emph{Ministerio de Ciencia e Innovaci\'on}}, project MTM2012-11686,  and thanks CWI, Amsterdam, for scientific support. Finally, some mathematical discussions with N. Sobrino-Coll are acknowledged.

\appendix

\section{R\'enyi-like functionals of Laguerre polynomials with large parameters} \label{A}

In this appendix the asymptotics ($\alpha \to \infty$) of some Rényi-like functionals of the Laguerre polynomials is given by means of the following theorem which has been recently found \cite{temme3} (see also \cite{temme1,temme2}). Herein, the involved parameters are just algebraic numbers without any quantum interpretation.


\begin{thm}\label{T1bis}
The R\'enyi-like functional of the Laguerre polynomials  $\mathcal{L}^{(\alpha)}_{m}(x)$ given by 
\begin{equation}
\label{eq:more01}
J_1(\sigma,\lambda,\kappa,m;\alpha) = \int\limits_{0}^{\infty}x^{\alpha+\sigma-1}e^{- \lambda x} \left|\mathcal{L}_{m}^{(\alpha)}(x)\right|^{\kappa}\,dx,
\end{equation}
(with $\sigma$ real, $0<\lambda \neq 1,\,\, \kappa >0$) has the following ($\alpha \to \infty$)-asymptotic behavior 
\begin{equation}
\label{eq:more13}
J_1(\sigma,\lambda,\kappa,m;\alpha)\sim\alpha^{\alpha+\sigma} e^{-\alpha}\lambda^{-\alpha-\sigma-\kappa m}\left\vert \lambda-1\right\vert^{\kappa m}
\sqrt{\frac{2\pi}{\alpha}}\frac{\alpha^{\kappa m}}{(m!)^\kappa}\sum_{j=0}^\infty \frac{D_{j}}{\alpha^j},
\end{equation}
with the first coefficients $D_0=1$ and 
\begin{equation}
\label{eq:more14}
\begin{array}{@{}r@{\;}c@{\;}l@{}}
D_1&=&\dsp{\frac{1}{12(\lambda-1)^2}\Bigl(1-12 \kappa m \sigma \lambda+6 \sigma^2 \lambda^2-12 \sigma^2 \lambda-6 \sigma \lambda^2+12 \sigma \lambda\ +}
\\
&&\quad\quad
6 \kappa^2 m^2+12 \kappa m \sigma-12 \kappa m^2 \lambda-12 \kappa m \lambda+6 \kappa m \lambda^2+\\
&&\quad\quad
6 \kappa m^2 \lambda^2+\lambda^2+6 \sigma^2-2 \lambda-6 \sigma+6 \kappa m^2\Bigr).
\end{array}
\end{equation}
\end{thm}
For the knowledge of the remaining coefficients and further details about the theorem, see \cite{temme3}.

\section{R\'enyi-like functionals of Gegenbauer polynomials with large parameters} \label{B}
In this appendix the asymptotics ($\alpha \to \infty$) of some Rényi-like functionals of Gegenbauer polynomials is given by means of the following theorem which has been recently found \cite{temme3} (see also \cite{temme1,temme2}). Herein, the involved parameters are just algebraic numbers without any quantum interpretation.

\begin{thm}\label{T3bis}
Let $a, b, c, d$, and $\kappa$ be positive real numbers, $c<d$, and $m$ a positive natural number. Then, the R\'enyi-like functional of the Gegenbauer polynomials $\mathcal{C}^{(\alpha)}_{m}(x)$ given by
\begin{equation}
\label{eq:intInl}
J_2(a,b,c,d,\kappa,m;\alpha) = \int_{-1}^{1} (1-x)^{c\alpha+a} (1+x)^{d\alpha+b}\left|\mathcal{C}^{(\alpha)}_{m}(x)\right|^{\kappa}\,dx 
\end{equation}
has the following asymptotics:
\begin{equation}
\label{eq:asymInl1}
J_2(a,b,c,d,\kappa,m;\alpha) \sim e^{-\alpha \phi(x_m)}\sqrt{\frac{2\pi}{\alpha}}\frac{2^{\kappa m}\bigl((\alpha)_m\bigr)^\kappa}{(m!)^\kappa} \sum_{k=0}^\infty\frac{D_k}{\alpha^k},\quad \alpha\to\infty
\end{equation}
where the coefficients $D_k$ do not depend on $\alpha$. The first coefficient is given by
\begin{equation}
\label{eq:Geg20}
D_0= a_1 \left(\frac{2c}{c+d}\right)^{a} \left(\frac{2d}{c+d}\right)^{b} \left(\frac{d-c}{c+d}\right)^{\kappa m},
\end{equation}
and the symbols $x_{m}=(d-c)/(d+c)$, $\phi(x_{m})=-c\log\frac{2c}{c+d}-d\log\frac{2d}{c+d}$ and $a_1=2\sqrt{\frac{cd}{(c+d)^{3}}}$. \\
Moreover, if $c=d$, the corresponding R\'enyi-like functional
\begin{equation}
\label{eq:asymInl2}
J_2(a,b,c,\kappa,m;\alpha)  = \int_{-1}^{1} (1-x)^{a} (1+x)^{b}e^{-\alpha\phi(x)}\left|\mathcal{C}^{(\alpha)}_{m}(x)\right|^{\kappa}\,dx,
\end{equation}
has the asymptotic behavior
\begin{equation}
\label{eq:Geg29}
J_2(a,b,c,\kappa,m;\alpha)  \sim \sqrt{\frac{\pi}{\alpha c}}\frac{(2\alpha)^m}{m!},\quad \alpha\to\infty.
\end{equation}

\end{thm}
Finally, the asymptotics of the R\'enyi-like functional with $c > d$ follows from the one with $c<d$ by interchanging $a$ and $b$ and $c$ and $d$. The case $c>d$ is useful for the determination of the R\'enyi entropy of the large dimensional hydrogenic states in momentum space with $q<1$. For further details of the theorem, see \cite{temme3}.


\section{Large-$D$ behavior of the angular R\'enyi factor $\Lambda_{l,\{\mu\}}(\Omega_{D-1})$}
\label{details}

Here we gather the necessary steps to obtain the final asymptotical expression (\ref{eq:rehabis}) for the angular R\'enyi factor $\Lambda_{l,\{\mu\}}(\Omega_{D-1})$ from Eq. (\ref{eq:reha2}).
First of all we have that since $\alpha_{j+1}=\alpha_j-\frac12$ one has that
\[\prod_{j=1}^{D-2}\frac{\Gamma(\alpha_j+q\mu_{j+1}+\frac12)}{\Gamma(\alpha_j+q\mu_j+1)}=\frac{\Gamma(\alpha_{D-2}+q\mu_{D-1}+\frac12)}{\Gamma(\alpha_1+q\mu_1+1)}=\frac{\Gamma(1+q\,\mu_{D-1})}{\Gamma(\frac D2+q\,l)}\]
Then, we can express the product
\[\prod_{j=1}^{D-2}4^{q(\alpha_j+\mu_j)}\frac{\Gamma(\alpha_j+\mu_{j+1})^{2q}}{\Gamma(2\alpha_j+\mu_{j+1}+\mu_j)^q}=
\prod_{j=1}^{D-2}4^{q(\alpha_j+\mu_j)}\frac{\Gamma(\alpha_j+\mu_{j+1})^{2q}}{\Gamma(2\alpha_j+2\mu_{j+1})^q}\prod_{j=1}^{D-2}\frac{\Gamma(2\alpha_j+2\mu_{j+1})^{q}}{\Gamma(2\alpha_j+\mu_{j+1}+\mu_j)^q},
\]
whose first factor can be rewritten as
\begin{eqnarray*}
&&\prod_{j=1}^{D-2}4^{q(\alpha_j+\mu_j)}\frac{\Gamma(\alpha_j+\mu_{j+1})^{2q}}{\Gamma(2\alpha_j+2\mu_{j+1})^q}=\prod_{j=1}^{D-2}4^{q(\mu_j-\mu_{j+1})}(2\sqrt \pi)^q\frac{\Gamma(\alpha_j+\mu_{j+1}+1)^{q}}{\Gamma(\alpha_j+\mu_{j+1}+\frac12)^q}(\alpha_j+\mu_{j+1})^{-q}\\\nonumber
&=& 4^{q(l-\mu_{D-1})}(2\sqrt \pi)^{q(D-2)}\frac{\Gamma(\frac D2+l)^q}{\Gamma(1+\mu_{D-1})^q}\prod_{j=1}^{D-2}\frac{\Gamma(\alpha_j+\mu_{j+1}+1)^q}{\Gamma(\alpha_j+\mu_{j}+1)^q}(\alpha_j+\mu_{j+1})^{-q}
\end{eqnarray*}
Taking into account these previous observations, Eq. (\ref{eq:reha2}) becomes 
\begin{eqnarray}
\Lambda_{l,\{\mu\}}(\Omega_{D-1}) &\sim& 2^{1-q} 4^{q(l-\mu_{D-1})} \pi^{1-q\frac D2}\mathcal M(D,q,\{\mu\})\,(\mathcal A(D,\{\mu\}))^q\times\\\nonumber
&& \frac{\Gamma(\frac D2+l)^q}{\Gamma(\frac D2+q\,l)}\frac{\Gamma(1+q\mu_{D-1})}{\Gamma(1+\mu_{D-1})^q}\prod_{j=1}^{D-2}\frac{\Gamma(2\alpha_j+2\mu_{j+1})^{q}}{\Gamma(2\alpha_j+\mu_{j+1}+\mu_j)^q}\frac{\Gamma(\alpha_j+\mu_{j+1}+1)^q}{\Gamma(\alpha_j+\mu_{j}+1)^q}
\end{eqnarray}
where
\[\mathcal M(D,q,\{\mu\})\equiv \prod_{j=1}^{D-2}\frac{\Gamma\big(q(\mu_j-\mu_{j+1})+\frac12\big)}{\Gamma\big(\mu_j-\mu_{j+1}+1\big)^q}\]
and
\[\mathcal A(D,\{\mu\})\equiv\prod_{j=1}^{D-2}(\alpha_j+\mu_j) (\alpha_j+\mu_{j+1})^{2(\mu_j-\mu_{j+1})-1}\]
Now, for convenience we introduce the notation 
\begin{eqnarray*}
\tilde{\mathcal E}(D,\{\mu\})&\equiv &\mathcal A(D,\{\mu\})\times\prod_{j=1}^{D-2}\frac{(2\alpha_j+2\mu_{j+1})_{\mu_j-\mu_{j+1}}^{-1}}{(\alpha_j+\mu_{j+1}+1)_{\mu_j-\mu_{j+1}}}\\
&=&\prod_{j=1}^{D-2} \frac{(\alpha_j+\mu_{j+1})^{2(\mu_j-\mu_{j+1})}}{(2\alpha_j+2\mu_{j+1})_{\mu_j-\mu_{j+1}}}\frac{1}{(\alpha_j+\mu_{j+1})_{\mu_j-\mu_{j+1}}},
\end{eqnarray*}
%
so that we have
\begin{eqnarray*}
\Lambda_{l,\{\mu\}}(\Omega_{D-1}) &\sim& \left(\pi 2^{1-q+2q(l-\mu_{D-1})}\frac{\Gamma(1+q\mu_{D-1})}{\Gamma(1+\mu_{D-1})^q} \right)\\\nonumber
&&\times\left(\mathcal M(D,q,\{\mu\})\,\frac{\tilde{\mathcal E}(D,\{\mu\})^q}{\pi^{q\frac D2}}\frac{\Gamma(\frac D2+l)^q}{\Gamma(\frac D2+q\,l)}\right)
\end{eqnarray*}
where the dominant factor is $\frac{\Gamma(\frac D2+l)^q}{\Gamma(\frac D2+q\,l)}\sim\Gamma(\frac D2)^{q-1}$. Moreover we can still simplify this expression by the use of the notation $\tilde{\mathcal M}$
\[\tilde{\mathcal M}(D,q,\{\mu\})\equiv 4^{q(l-\mu_{D-1})}\,\pi^{1-\frac{D}{2}}\mathcal M(D,q,\{\mu\})\]
to have $\tilde{\mathcal M}\equiv1$ for all configurations with $\mu_1=\mu_2=...=\mu_{D-1}$. Thus, we can finally obtain the searched expression (\ref{eq:rehabis}); namely, 
\begin{eqnarray*}
\Lambda_{l,\{\mu\}}(\Omega_{D-1}) &\sim& \left( 2^{1-q}\frac{\Gamma(1+q\mu_{D-1})}{\Gamma(1+\mu_{D-1})^q} \right)\\\nonumber
&&\times\left(\tilde{\mathcal M}(D,q,\{\mu\})\tilde{\mathcal E}(D,\{\mu\})^q \pi^{\frac D2(1-q)}\frac{\Gamma(\frac D2+l)^q}{\Gamma(\frac D2+q\,l)}\right)
\end{eqnarray*}
This expression together with (\ref{eq:renyihyd3}) allow us to determine the dominant term of the angular R\'enyi entropy for fixed $l$ as
\begin{equation}
R_{q}[\mathcal{Y}_{l,\{\mu\}}] \sim -\log\left(\Gamma\left(\frac D2\right)\right)+\frac D2\log\pi+\frac 1{1-q}\log\left(\tilde{\mathcal E}(D,\{\mu\})^q\tilde{\mathcal M}(D,q,\{\mu\})\right)
\end{equation}
(where the third term vanishes for $\mu_1=\mu_2=...=\mu_{D-1}$) which correponds to expression (\ref{eq:Rq_ang}) with the values of $\tilde{\mathcal M}(D,q,\{\mu\})$ and $\tilde{\mathcal E}(D,\{\mu\})$ given above.

\end{document}